\documentclass[preprint2]{aastex}

\usepackage{amsmath}
\usepackage{natbib}

\newcommand{\msun}{{\rm M}_{\sun}}

\newcommand{\rsun}{{\rm R}_{\sun}}

\newcommand{\rstar}{{\rm R}_{\star}}

\newcommand{\lumsun}{{\rm L}_{\sun}}

\newbox\grsign \setbox\grsign=\hbox{$>$}
\newdimen\grdimen \grdimen=\ht\grsign
\newbox\laxbox \newbox\gaxbox
\setbox\gaxbox=\hbox{\raise.5ex\hbox{$>$}\llap
     {\lower.5ex\hbox{$\sim$}}}\ht1=\grdimen\dp1=0pt
\setbox\laxbox=\hbox{\raise.5ex\hbox{$<$}\llap
     {\lower.5ex\hbox{$\sim$}}}\ht2=\grdimen\dp2=0pt

\begin{document}

\title{Accretion disks around massive stars: \\
 Hydrodynamic structure, stability and dust sublimation }

\author{Bhargav Vaidya\altaffilmark{\textit{a}}, Christian Fendt and Henrik Beuther}

\affil{Max Planck Institute for Astronomy, K\"onigstuhl 17,
D-69117 Heidelberg, Germany} \email{vaidya@mpia.de , fendt@mpia.de
, beuther@mpia.de}

\altaffiltext{\textit{a}}{Member of the \textit{International Max
Planck Research School for Astronomy and Cosmic Physics at
University of Heidelberg}, IMPRS-HD}

\begin{abstract}

We investigate the structure of accretion disks around massive
protostar applying steady state models of thin disks.
The thin disk equations are solved with proper opacities for dust
and gas taking into account the huge temperature variation along
the disk.
We explore a wide parameter range concerning stellar mass,
accretion rate, and viscosity parameter $\alpha$.
The most essential finding is a very high temperature of the inner
disk. For e.g.\ a $10\,\msun$ protostar and an accretion rate of
$\sim 10^{-4}\,\msun {\rm yr}^{-1}$, the disk midplane temperature
may reach almost $10^5\,$K. The disk luminosity in this case is
about $10^4\,\lumsun$ and, thus, potentially higher than that of a
massive protostar.
We motivate our disk model with similarly hot disks around compact
stars.
We calculate a dust sublimation radius by turbulent disk
self-heating of more than $10\,{\rm AU}$, a radius, which is 3 times
larger than caused by stellar irradiation. We discuss implications
of this result on the flashlight effect and the consequences for
the radiation pressure of the central star.
In difference to disks around low mass protostars our models
suggest rather high values for the disk turbulence parameter
$\alpha \leq 1$. However, disk stability to fragmentation due to
thermal effects and gravitational instability would require a
lower $\alpha$ value. For $\alpha = 0.1$ we find stable disks out
to $80\,{\rm AU}$.
Essentially, our model allows to compare the outer disk to some of
the observed massive protostellar disk sources, and from that,
extrapolate on the disk structure close to the star which is yet
impossible to observe.
\end{abstract}

\keywords{accretion, accretion disks - hydrodynamics - methods :
analytical - stars: formation - turbulence }

\section{Introduction}
Massive stars play a vital role in order to understand the
dynamical evolution of clusters in which they are the major source
of heavy elements and UV radiation. During their short life time,
they impact their surrounding by a number of physical processes
such as jet like outflows, strong winds, photo evaporation,
expanding HII regions and eventually supernova explosions
\citep{2007ARA&A..45..481Z,2007arXiv0712.1109B}. Collimated
molecular outflows and jets from young massive protostars have
been observed \citep{2002A&A...383..892B, 2005ccsf.conf..105B},
although the launching mechanism for outflows have been mainly
investigated for low mass protostars so far (see e.g.
\citealt{2002ApJ...581..988C, 2007prpl.conf..277P,
2009ApJ...692..346F}).

Understanding massive star formation has been a very active field
of research for both the observers and the theorists. The basic
differences of high and the low mass star formation process are
that of timescales, energy scales, mass flow rates and high
luminosity of the central star. Low mass stars have well defined
phases in their formation process and they only start burning
hydrogen after accretion of all matter is done
\citep{2005fost.book.....S}. In contrast, massive stars have very
short Kelvin Helmholtz time of $\sim 10^4 - 10^5\,{\rm yr}$, and
thus start burning hydrogen even when they are still accreting.
One of the problems in the formation of massive stars is that, at
some point in time, the strong radiation pressure from the
luminous central star may exceed the Eddington limit in spherical
symmetry and does not allow the matter to accrete anymore. This
seems to limit the mass of the star to be formed via the spherical
accretion to about $20\,\msun$ \citep{1987ApJ...319..850W}.

However, disk accretion may add dynamical pressure to the
accreting matter which may help overcoming the stellar radiation
pressure. Thus, the star formation scenario for low and high mass
stars could in principle be the same (e.g.
\citealt{2007ApJ...666..976K}), suggesting that high mass star
formation as a scaled up version of the low mass star formation
process. With our paper we aim to construct a {\em global model}
(length scales of $\sim 0.1 {\rm AU}$ to $100 {\rm AU}$) of the
accretion disk around massive young stars. By fitting the outer
disk structure to the observations, this will allow us to
investigate, in follow up projects, the feasibility of certain
physical processes in inner disk which could be essential for the
pre-stellar evolution as angular momentum transport or processes
responsible for launching of outflows.

Our paper relies on one essential assumption - that is the disk
accretion rate. Both estimates following the observed {\em
outflow} rates and the time scale of mass aggregation suggest
accretion rates in the range of $10^{-3} - 10^{-4} {\msun} {\rm
yr}^{-1}$
\citep{2002A&A...383..892B,2005IAUS..227..135Z,2009arXiv0901.2053G}.
However, one of the aims of our study is to understand whether
such values are compatible with theoretical accretion disk models.

In the following section of this paper, we discuss about the
modelling of the disk structure using the standard disk equations,
in \S 3 we describe the disk opacity model which has contribution
from both, dust and gas and in \S 4 we present our results and
give estimates of the sublimation radius and then in \S 5 discuss
about the stability of the disk and how the viscosity parameter
$\alpha$ has an effect on it. Also in \S 6 we compare our results
with the recent observations of disks in massive young protostars.
Finally, in \S 7 we present our conclusions from this simple
approach to study disks around massive stars.

\section{Accretion Disk Model}

%
In order to study the {\em global} disk structure around massive
stars, we apply the thin disk model
\citep{1973A&A....24..337S,1989Icar...82..225M,
1991ApJ...375..740R, 1992apa..book.....F,1993Icar..106...77S,
1994ApJ...427..987B} with appropriate modifications. The essential
part of our model is to take into account the proper gas and dust
opacities for a huge temperature regime as indicated for high
accretion rates on to massive stars. The original standard disk
model by \cite{1973A&A....24..337S} relates to solving the
stationary state hydrodynamic equations for the thin disk assuming
certain opacity laws for different regions in the disk. The disk
is classified to have three regions (A,B, and C) depending upon
the opacity and the pressure. The region C is the outermost region
where the gas pressure is much greater than the radiation pressure
in the disk and free-free opacity dominates in this region,
whereas region B is the one in which gas pressure is still greater
but the Thomson scattering is now dominating over Kramer's
opacity. The region A is the one which is closest to the central
object and here the radiation pressure in the disk is much greater
than the gas pressure. The thin disk model was extended to include
dust opacities and has been applied to disks around low mass stars
and solar nebula.
\citep{1991ApJ...375..740R,1993Icar..106...77S,1998Icar..132..100S,2002MNRAS.332..485D}.
The application of such a thin disk model to massive stars is a
simple approach to understand the dynamics of the disk especially
in the inner most region which can not be resolved with the help
of present day telescopes. We look for steady state solution of
the thin disk model using the hydrodynamic equations of the thin
disk (e.g.\ \citealt{1992apa..book.....F}). The vertically
averaged surface density $\Sigma$ is related to the mass density
$\rho$ and the scale height H.
\begin{equation}\label{eq:1}
   \Sigma = 2H\rho
\end{equation}
The scale height is estimated using the approximation for thin
disk
\begin{equation}\label{eq:2}
    H = \sqrt{2}\left(\frac{c_{s}}{\Omega}\right)
\end{equation}
where $\Omega$ is the Keplerian angular velocity and $c_{s}$, the
sound speed is defined in terms of temperature and other physical
constants like Boltzmann constant $k$, proton mass $m_{p}$ and the
mean molecular weight $\mu$.
\begin{equation}\label{eq:3}
    {c_{s}}^{2} = \left(\frac{k}{\mu m_{p}}\right) T
\end{equation}
The energy balance equation which implies that the amount of
energy radiated from the disk is equal to the amount of energy
produced by viscous heating.
\begin{equation}\label{eq:4}
    \left(\frac{16\sigma}{3\Sigma\kappa}\right)T^{4} =
    \left(\frac{9}{4}\right)\nu\Sigma\Omega^{2}
\end{equation}
The viscosity $\nu$ is defined by introducing the parameter
$\alpha$.
\begin{equation}\label{eq:5}
    \nu = \alpha c_{s} H
\end{equation}
The Rosseland mean opacity $\kappa$ is related to the mass density
and the temperature in form of power law with $\kappa_{0}$ as
constant.
\begin{equation}\label{eq:6}
    \kappa = \kappa_{0} \rho^{m} T^{n}
\end{equation}
The condition for the conservation of angular momentum in steady
state can also be derived.
\begin{equation}\label{eq:7}
    \Sigma\nu = \frac{\dot{M}}{3\pi}\left[1 -
    \left(\frac{R_{*}}{r}\right)^\frac{1}{2}\right]
\end{equation}

The dimensionless viscosity factor $\alpha$ is introduced
basically to parameterize viscosity by assuming that is provided
by turbulent motions present in the disk. The physical
significance of equation~\eqref{eq:5} is that the largest size of
turbulent eddy that can be sustained in the disk is of size equal
to the disk scale height and maximum speed of the turbulent motion
is that of sound speed, this implies values of $\alpha \lesssim
1$. In order to solve the above equations consistently, one needs
the form of opacity. For the opacity of the form of a power law
given by equation~\eqref{eq:6}, one gets power law solutions of
all the dynamical quantities in the disk. For instance, in case of
the form of Kramer opacity law ($\kappa = 6.6\times10^{-22}
\rho^{1}T^{-\frac{7}{2}}$) used by \citet{1973A&A....24..337S},
the central temperature in the disk around a typical massive star
in the region where $ P_{\rm gas} > P_{\rm rad}$  and $
\sigma_{\rm ff} > \sigma_{\rm t}$ is
{\setlength\arraycolsep{2pt}
\begin{eqnarray}
\label{eq:8} T_{\rm c}(r) = 1.74\times10^{4}\,{\rm K}\,
\left(\alpha \right)^{-1/5} \left(\frac{\dot{M}}{10^{-4}
\msun\,{\rm yr}^{-1}}\right)^{3/10}
\nonumber \\
\left(\frac{M}{10\,\msun}\right)^{1/4}
\left(\frac{r}{10\,\rstar}\right)^{-3/4} \left(\frac{
\rstar}{6\,\rsun}\right)^{-3/4}
\end{eqnarray}}
where $\sigma_{\rm ff}$ is the free free opacity and $\sigma_{\rm
t}$ is the opacity due to electron scattering.


\section{Disk opacities}

The work done by \citet{1973A&A....24..337S} was mainly focused on
hot accretion disks around black holes, the contribution to the
opacity in their case was mainly from the electron scattering or
free-free emission depending upon the optical depth in the disk.
However in disks around massive young stars, contribution to
opacity also comes from dust present beyond the sublimation
radius, many molecular and atomic lines and other scattering
processes. Since massive stars have large radiation fields that
heavily affect the huge amount of dust present in the envelope and
also the gas that is present in the inner part, one has to take
into account a proper contribution of opacities from dust and gas
to have a consistent accretion disk model.

In our model the dust opacity does not have any explicit
dependence on frequency, so the effective value of the opacity
\begin{equation}\label{eq:12}
     \kappa_{\rm eff} =  \frac{\int \kappa_{\nu}F_{\nu} d{\nu}}{\int F_{\nu} d{\nu
     }}
\end{equation}
simplifies to $\kappa_{\rm eff} = \kappa_{\rm \nu}$


In our work, we follow that the matter is accreted via a disk
investigating the possible location of the dust sublimation
radius. We expect that the active disks will be dominating in
destroying the dust in the disk at a much further radius then it
would have been destroyed by the heating from star. This would
result in lowering the radiation pressure on the dust and allowing
the infall of matter on the central star.

Apart from some metal silicates all the dust grains that are
present sublimates at around $1500\, {\rm K}$. Since at this
temperature, the gas and the dust opacities vary substantially, it
is essential for calculating radiation pressure to consider the
proper opacity.

\subsection{Dust opacities}
There are many factors one has to take into account for getting a
consistent model for dust opacity such as the size of the grains,
distribution of grains in the disk and the coagulation of dust
grains. Also the temperature of the disk varies over a large range
and this would clearly affect the composition of the dust and
alter the gas to dust ratio typically $\sim 100$ in the disk. This
radial variation in the gas to dust ratio will not only modify the
gas pressure along the disk but also the other dynamical disk
quantities. The usual assumption in many of the dust models is
that the dust and gas are well coupled and also the gas to dust
ratio is taken to be 100. In general, opacity has a typical
dependence on temperature and density. (e.g.\
\citealt{1989Icar...82..225M,1991ApJ...375..740R,1994ApJ...427..987B,1994A&A...291..943O,2000A&A...358..651H,2003A&A...410..611S}).

In the regime which is dominated by dust  ($r\gg R_{*}$), the
opacity depends very weakly on density. In the present work we
apply for the dust dominated region the model proposed by
\citet{1991ApJ...375..740R} in which the Rosseland mean opacities
have been given in terms of the power laws of the form given in
equation~\eqref{eq:6}, in different regimes that have been
distinguished on the basis of temperature and mass density (see
Table~\ref{tab1}). Using this power law form of opacity,
equations~\eqref{eq:1} - \eqref{eq:6} can be reduced to a form
representing the viscosity as a function depending on surface
density and the radial distance as
\begin{equation}\label{eq:13}
    \nu = C \Sigma^{p} r^{q}
\end{equation}
This form of viscosity can be then substituted in
equation~\eqref{eq:7} so that all the dynamical quantities namely
the surface density, mass density, central temperature, scale
height of the disk can be expressed as power laws of mass,
accretion rate, radial distance and the viscosity parameter
$\alpha$.
For the parameters of a typical massive star these quantities (Q)
can be written
{\setlength\arraycolsep{2pt}
\begin{eqnarray}\label{eq:14}
Q =
C_{\textit{i}}\left(\frac{M_{*}}{\msun}\right)^{\gamma_{\textit{i}}}
\left(\frac{r}{R_{\textit{i}}}\right)^{\epsilon_{\textit{i}}}\alpha^{\delta_{\textit{i}}}
\nonumber \\
\left(\frac{\dot{M}}{10^{-3}\msun {\rm
yr}^{-1}}\right)^{\beta_{\textit{i}}}
\end{eqnarray}
e.g.\
\citep{1991ApJ...375..740R,1993Icar..106...77S,2002MNRAS.332..485D}.
The index \textit{i} denotes the opacity regime as given in the
Table ~\ref{tab1}. ($\textit{i} = 1,2...6$), $C_{\textit{i}}$ are
constants and $R_{\textit{i}}$ is chosen as a typical radius for
that regime. The dust model taken into account does not consider
explicit wavelength dependence of opacity.
\begin{table*}
\scriptsize
\begin{center}
\caption{Different regimes of the opacity taken from
\citet{1991ApJ...375..740R} with the conditions for the change
over and description of the main components of dust in each regime
} \vspace{0.6cm} \label{tab1}
\begin{tabular}{c c c c}

  \hline\hline
  \noalign{\smallskip}

  Regime & $\kappa_{\textit{i}}$ & Condition & Description \\

  \hline

   \noalign{\smallskip}

  1 & $\left(2\times10^{-4}\right) \rho^{0} T^{2}$ & $T \leq 150\,{\rm K}$ & dominated by Water and Ice grains \\

  2 & $\left(1.15\times10^{18}\right) \rho^{0} T^{-8}$ & $T \leq 180\,{\rm K}$ & Sublimation of Ice grains  \\

  3 & $\left(2.13\times10^{-2}\right) \rho^{0} T^{0.75}$ & $T \leq 1380\, {\rm K} \left(\frac{\rho}{10^{-8}}\right)^{1/50}$ & mainly consists of Iron and Silicate grains \\

  4 & $\left(1.57\times10^{60}\right) \rho^{3/8} T^{-18}$ & $T \leq 1890\, {\rm K} \left(\frac{\rho}{10^{-8}}\right)^{1/48}$ & Sublimation of refractory grains  \\

  5 & $\left(1.6\times10^{-2}\right) \rho^{0} T^{0}$ & $T \leq 2620\, {\rm K} \left(\frac{\rho}{10^{-8}}\right)^{2/27}$ & Molecular and atomic lines contribute \\

  6 &  $\left(2\times10^{34}\right) \rho^{2/3} T^{-9}$ & $T \leq 3200\, {\rm K}$ & Molecular and atomic lines contribute   \\

   \noalign{\smallskip}
  \hline\hline
   \noalign{\smallskip}

\end{tabular}

\end{center}

\end{table*}


\subsection{Gas opacities}

\def\thefootnote{\fnsymbol{ctr}}
\long\def\symbolfootnote[#1]#2{\begingroup%
\def\thefootnote{\fnsymbol{footnote}}\footnote[#1]{#2}\endgroup}
The opacity of gas and dust have been estimated consistently. These opacities are usually
listed in form of a table in two parameters space, temperature (T)
and parameter (R) which is dependent on density in a following
manner
\begin{equation}\label{eq:9}
  R = \frac {\rho}{\left(T_{6}\right)^{3}}
\end{equation}
\citep{1996ApJ...464..943I,2005ApJ...623..585F} where $\rho$ is
the density and $T_{6}$ is the temperature normalized to $10^{6}\,
{\rm K}$. These opacities also show some dependence on metallicity
as well.

In a regime where dust begins to sublimate and molecules
start to form, there is a sharp decrease in the opacity. In this regime,
dependence of opacity on density is strong, unlike the opacity due
to dust, and in general it is very difficult to get the true value of
the opacity in this region so usually linear interpolation is used
\citep{2003A&A...410..611S}. Due to
steep turn over of the opacity gradient with the temperature, numerical difficulties arise when the
dynamical quantities are estimated for this region.
In the present work, we take into account the opacity
table\symbolfootnote[2]{http://webs.wichita.edu/physics/opacity}
by \citet{2005ApJ...623..585F}, which has the temperature range of
our relevance ($500\,{\rm K} $ - $10^{4.5}\,{\rm K} $). For the
present purpose we choose a table with hydrogen fraction  X = 0.94
and metallicity Z = 0.06. In fact, we also in the inner most
region obtain higher temperatures and so we apply in continuation
the OPAL opacity
tables\symbolfootnote[3]{http://www-phys.llnl.gov/Research/OPAL/opal.html}
for higher temperatures ($\geq 10^{4.15}\, {\rm K}$)
\citep{1996ApJ...464..943I}. We create equispace grid of two
parameters log(T) and log(R) as defined above and put the values
from the table in the grid and fit a 2D cubic spline on it to get
the interpolated values of opacity for any given temperature and
density, but as the regime where the gas and dust coexists is
difficult to model numerically as the opacity do not converge to a
unique value, we make a linear approximation as zeroth order
approach and compare with the analytical models present. (e.g.\
\citealt{1991ApJ...375..740R,1994ApJ...427..987B}).

%


\section{Results and Discussion}
In the following we present results of applying above standard
models of disk accretion and opacities to massive star formation.
We first discuss the outer dust dominated disk and then the inner
gaseous disk.


\subsection{Outer disk structure : Radial Profiles of Dynamical Quantities}
The main contribution of opacity in the outer disk is from dust.
The numerical values for the dynamical quantities expressed as
given by equation~\eqref{eq:14} in different opacity regimes are
listed in Table~\ref{tab2}. The radial profile of the disk
midplane temperature, surface density, mass density and scale
height for a high mass star is shown in Fig.~\ref{fig:radprofdust}
for a $10 {\rm M_{\sun}}$ star with accretion rate of the order of
$4.2\times10^{-4}\,\rm {M_{\sun}}{\rm yr^{-1}}$ and the a
viscosity parameter $\alpha = 1$. With these parameters, the
central temperature reaches $1500\, {\rm K} $ around 12\,AU due to
viscous heating, which causes the dust to sublimate. Figure
~\ref{fig:radprofdust} shows two curves for each quantity. The
solid curve is obtained by implementing the opacity power laws as
given by \citet{1991ApJ...375..740R}, whereas the dashed line is
obtained by using the opacities by \cite{1993Icar..106...77S} and
extrapolating it to higher mass stars.

There are few kinks seen in the plot, which are due to the fact
that the dust opacity model used comprises of different regimes
and in each of these regimes have a different form of power law
(see Table ~\ref{tab1}). These regimes are connected using the
standard procedure \citep{1985prpl.conf..981L} of equating the
opacity in two consecutive regimes, $\kappa_{\textit{i}} =
\kappa_{\textit{i}+1}$.

Since formation of massive stars involves large accretion rates
the surface density and the mass density in the disk is higher
than the lower mass counterparts, as more matter will be injected
in the inner region. We find the scale height ratio obtained from
the present model is approximately constant in the outer disk,
with a radial dependence $\propto r^{0.28}$ and consistent with
the thin disk approximation.

A least square linear fit of dynamical quantities gives for the
temperature profile a power for the radial distance as -0.45 and
for density as -2.35. The surface density has a rather flat
profile as the best fit gives the dependence of the form $\propto
r^{-1.1}$. These profiles are similar to those obtained by
\citet{1991ApJ...375..740R} but here applied for a high mass star
with high accretion rate.

The parameter values discussed above were chosen i) to have most
part of the inner 100 AU disk to be gravitationally stable and
also ii) to ensure that the dust in the disk sublimates at a
distance substantially further than caused by sublimation from
stellar irradiation. (see below)
\begin{table}
\scriptsize
\begin{center}
\caption{The various dynamical quantities for the massive star
using the interpolated Rosseland mean dust opacity model given by
\citet{1991ApJ...375..740R}.(see equation~\eqref{eq:14} and
Table~\ref{tab1})}\vspace{0.6cm}\label{tab2}
\begin{tabular}{c c c c c}
  \hline
  \hline
  & \multicolumn{4}{c}{Regime 1 $R_{1} = 400 AU$} \\
  \hline

  \noalign{\smallskip}
  Q  & $T[{\rm K}]$ & $Height [{\rm cm}]$ & $\rho [{\rm g}{\rm cm}^{-3}]$ & $\Sigma [{\rm g}{\rm cm}^{-2}]$ \\
  \noalign{\smallskip}
  \hline
  \noalign{\smallskip}
  $C_{\textit{i}}$  & 126.3 & $9.9\times10^{14}$  & $1.3\times10^{-13}$  & 258.8\\
\hline
\end{tabular}
\vspace{0.5cm}
\\
\begin{tabular}{c c c c c }
\hline \hline
& \multicolumn{4}{c}{Regime 2 $R_{2} = 250 AU$} \\
\hline
\noalign{\smallskip}
Q  & $T[{\rm K}]$ & $Height [{\rm cm}]$ & $\rho [{\rm g}{\rm cm}^{-3}]$ & $\Sigma [{\rm g}{\rm cm}^{-2}]$ \\
\noalign{\smallskip}
  \hline
  \noalign{\smallskip}
$C_{\textit{i}}$  & 169.8 & $5.7\times10^{14}$  & $3.4\times10^{-13}$  & 393.1\\
\hline
\end{tabular}
\vspace{0.5cm}
\\
\begin{tabular}{c c c c c}
 \hline
 \hline
  & \multicolumn{4}{c}{Regime 3 $R_{3} = 50 AU$} \\
  \hline
  \noalign{\smallskip}
  Q  & $T[{\rm K}]$ & $Height [{\rm cm}]$ & $\rho [{\rm g}{\rm cm}^{-3}]$ & $\Sigma [{\rm g}{\rm cm}^{-2}]$ \\
  \noalign{\smallskip}
  \hline
  \noalign{\smallskip}
  $C_{\textit{i}}$  & 814.5 & $1.12\times10^{14}$   & $4.1\times10^{-12}$  & 907.6  \\
  \hline
\end{tabular}
\vspace{0.5cm}
\\
\begin{tabular}{c c c c c}
 \hline
 \hline
  & \multicolumn{4}{c}{Regime 4 $R_{4} = 10 AU$} \\
  \hline
  \noalign{\smallskip}
  Q  & $T[{\rm K}]$ & $Height [{\rm cm}]$ & $\rho [{\rm g}{\rm cm}^{-3}]$ & $\Sigma [{\rm g}{\rm cm}^{-2}]$ \\
  \noalign{\smallskip}
  \hline
  \noalign{\smallskip}
  $C_{\textit{i}}$  & 1613.2 & $1.4\times10^{13}$   &  $1.8\times10^{-10}$ & 5126.4  \\
  \hline
\end{tabular}
\vspace{0.5cm}
\\
\begin{tabular}{c c c c c}
 \hline
 \hline
  & \multicolumn{4}{c}{Regime 5 $R_{5} = 6 AU$} \\
  \hline
  \noalign{\smallskip}
  Q  & $T[{\rm K}]$ & $Height [{\rm cm}]$ & $\rho [{\rm g}{\rm cm}^{-3}]$ & $\Sigma [{\rm g}{\rm cm}^{-2}]$ \\
 \noalign{\smallskip}
  \hline
  \noalign{\smallskip}
  $C_{\textit{i}}$  & 1918.3 & $7.1\times10^{12}$   &  $6.5\times10^{-10}$ & 9282.8  \\
  \hline
\end{tabular}
\vspace{0.5cm}
\\
\begin{tabular}{c c c c c}
 \hline
 \hline
  & \multicolumn{4}{c}{Regime 6 $R_{6} = 3 AU$} \\
  \hline
  \noalign{\smallskip}
  Q  & $T[{\rm K}]$ & $Height [{\rm cm}]$ & $\rho [{\rm g}{\rm cm}^{-3}]$ & $\Sigma [{\rm g}{\rm cm}^{-2}]$ \\
  \noalign{\smallskip}
  \hline
  \noalign{\smallskip}
  $C_{\textit{i}}$  & 2767.9 & $3.0\times10^{12}$    & $3.0\times10^{-9}$  & $1.8\times10^{4}$  \\
  \hline
\end{tabular}
\end{center}
\end{table}


\subsection{Sublimation of dust}
Most of the dust grains sublimate when the temperature in the disk
reaches the critical value of 1500 K. The radius of the disk at
which this value of critical temperature is reached is the dust
sublimation radius. The disk can be heated by various processes
such as viscous heating, stellar irradiation, convection, cosmic
rays \citep{1998ApJ...500..411D}. However, in the present work we
estimate the dust sublimation radius via two main processes namely
viscous heating and stellar irradiation (see Table~\ref{tab3}).
\subsubsection{Disk Self-Sublimation of dust}
For our massive stellar disk model, we define that radius where
the disk temperature reaches 1500 K as disk self-sublimation
radius. Essentially, the disk temperature is determined by
$\dot{M}$, while the radial velocity is governed by $\alpha$. The
higher the accretion rate, more the gravitational potential energy
from the accreted mass is converted to thermal energy in the inner
region, thus increasing the disk temperature. High mass accretion
rates  $\sim 10^{-3} - 10^{-4}\,{\rm \msun} {\rm yr}^{-1}$ are
indirectly related to the formation of massive stars, indicating
hotter disks in massive stars.

The dust sublimation radius can be estimated from the temperature
profile in regime 4 (see Table~\ref{tab2}) as the critical value
of 1500 K is obtained in this regime,
{\setlength\arraycolsep{2pt}
\begin{eqnarray}\label{eq:15}
T =  1.6\times10^{3}\, {\rm K}\,\alpha^{-0.058}
\left(\frac{\dot{M}}{10^{-3}\msun {\rm yr}^{-1}}\right)^{0.101}
\nonumber\\
\left(\frac{M}{\msun}\right)^{0.080} \left(\frac{r}{10 \rm
{AU}}\right)^{-0.239}.
\end{eqnarray}
Thus the dust self-sublimation radius is
{\setlength\arraycolsep{2pt}
\begin{eqnarray}\label{eq:16}
R_{\rm sub,disk} = 11\,{\rm AU}\, \alpha^{-0.242}
\left(\frac{\dot{M}}{10^{-4}\msun {\rm yr}^{-1}}\right)^{0.422}
\nonumber\\
\left(\frac{M}{10\msun}\right)^{0.3347} \left(\frac{T}{1500 {\rm
K}}\right)^{-4.184}.
\end{eqnarray}
\subsubsection{Dust Sublimation by Stellar radiation}
The second process of disk heating is by radiation from the
central star. The dust sublimation radius due to the absorption of
UV radiation from the star by the disk is dependent on the
effective temperature and the radius of the star or equivalently
to the luminosity of the star. We can estimate the relation
between the temperature of the disk and the temperature of the
star just by equating the flux from the star that is absorbed by
the disk to the flux emitted by it considering it as black body.
The disk is also assumed to be locally isothermal, which is an
appropriate assumption for optically thick
disks
The mid plane temperature of the disk due to heating from central
star can be estimated,
\begin{equation}\label{eq:17}
    T_{d} = \left({\frac{\theta \psi_{s}}{2\psi_{i}}}\right)^{1/4}\left({\frac{R_{*}}{r}}\right)^{1/2} T_{*}
\end{equation}
(e.g.\ \citealt{2001ApJ...560..957D}) where $\theta$ is the angle
with which the radiant flux is incident on the flaring disk,
$\psi_{s}$ denotes the fraction of flux that is absorbed by the
interior and $\psi_{i}$ is correction factor which accounts for
the fact that the disk interior is not fully optically thick for
its emission. The small correction due to back-warming in the disk
is neglected. This correction mainly depends on the underlying
dust properties \citep{2002ApJ...579..694M}. For a dust
sublimation temperature of $\sim 1500\, {\rm K} $, we can estimate
the sublimation radius due to absorption of radiation from the
above equation~\eqref{eq:17} noted as Measure A whereas Measure B
we denote as the sublimation radius calculated using the standard
formula given by \cite{2002ApJ...579..694M}. In Table~\ref{tab3}
we show the typical values of the sublimation radius from these
measures. For a typical B2 star of mass = 10$\msun$ and $T_{\rm
eff} = 22000\, {\rm K}$ \citep{1992adps.book.....L} with ZAMS
value for the stellar radius as 6 $\rsun$, one gets a dust
sublimation radius due to heating from stellar radiation of $
R_{\rm {sub,*}} \sim 4\, {\rm AU}$ for measure A and $\sim 3\,{\rm
AU}$ for measure B. This value may increase by a factor of 2
considering the bloating star model \citep{2008ASPC..387..255H},
where the star bloats up to $100\rsun$ and the effective
temperature reduces to $5000\,{\rm K}$.
\\
In Table~\ref{tab3}, we compare for different stellar mass with
typical order of mass accretion rate, the sublimation radius due
to self heating from the disk and that due to heating from the
star. This values are estimated for two different values of
$\alpha = 0.1, 1$. It is evident from Table~\ref{tab3} that the
dust sublimation radius due to heating from disk is a weak
function of $\alpha$. Table~\ref{tab3} further indicates a ratio $
R_{\rm {sub,disk}}/R_{\rm {sub,*}} \sim 3$ as a good estimate over
a wide range of mass accretion rates and stellar mass.

The value of $\alpha$ is related to the radial transport of matter
in the disk, thus the mass accretion rate is a function of
$\alpha$. For Table~\ref{tab3}, the values of the sublimation
radius is estimated using eq~\eqref{eq:15} assuming the same mass
accretion rate for different $\alpha$ values. These are
representative values of sublimation radius for typical mass
accretion rates found in massive star forming regions. Using these
values we optimize the ratio of the self sublimation radius of the
disk to that caused by heating from stellar luminosity is around
three. With this assumption, we obtain a relation of $\alpha$ and
$\dot{M}$ for a particular stellar mass. This is obtained by
setting $T = 1500\, {\rm K}$ in the equation~\eqref{eq:15} and the
radial distance as $3\times R_{\rm sub,*}$.

This relation can be used to have constraints on the mass
accretion rates for a particular stellar mass. The plot for mass
accretion rates with $\alpha$ is shown in Fig.~\ref{fig:mdotalpha}
for typical OB type stars. If one considers the value of $\alpha$
as fixed to 1, then from the Fig.~\ref{fig:mdotalpha} for the $10
M_\sun$ star one gets the accretion rate of ${4.2\times10^{-4}
M_\sun {\rm yr}^{-1}}$, which implies that for this order of mass
accretion rate the dust in the disk will sublimate at a distance
of $3\times R_{{\rm sub},*}$. This clarifies that for the typical
high accretion rates required for formation of high mass stars,
the heating in the disk is very efficient to sublimate most of the
dust grains in the midplane before the radiation could have any
major effect on them.

The implication of this result is profound, as it demonstrates
that viscous heating of the disk is the dominant mechanism in the
midplane for sublimation of the dust. The self-sublimation in the
turbulent massive disk sublimates most of the dust grains well
before the stellar radiation could affect them. Essentially, this
implies that more matter (in form of gas) can reach closer to the
central star.

However, stellar radiation  can have significant effects on the
surface layer of the disk. The radial distance beyond which the
stellar irradiation will dominate can be estimated by equating the
flux from central star to the flux emitted from the disk. The flux
from the disk is dependent on the effective surface temperature,
which can be substantially lower than the midplane temperature
depending on the optical depth. (see \S 4.3)

\begin{table*}
\scriptsize
\begin{center}
\caption{Comparison of the sublimation radius due to heating in
disk [equations~\eqref{eq:15} \& \eqref{eq:16}] and due to heating
from the star [equation~\eqref{eq:17}]. The temperature of the
star and its radius are taken from \citet{1992adps.book.....L}.
(see \S 4.2 for details)} \vspace{0.6cm} \label{tab3}
\begin{tabular} {c c c c c c c c }

  \hline\hline
 \noalign{\smallskip}

  Typical $\dot{M}$ [$\msun$ $\rm{yr}^{-1}$]& Spectral Type & Stellar Mass ($\msun$) & \multicolumn{2}{c}{$R_{\rm sub,*} [{\rm AU}]$} & \multicolumn{2}{c}{$R_{\rm sub,disk}$ [{\rm AU}]}   \\
  \noalign{\smallskip}
  \hline
  \noalign{\smallskip}
   & & & Measure A  & Measure B &     $\alpha = 0.1$ & $\alpha = 1$ \\

   \hline
   \noalign{\smallskip}

   $10^{-5}$& ${\rm B5}$ & 5.9 & 1.4 & 1.0 & 6.1 & 3.5 \\

   $10^{-4}$& ${\rm B2}$ & 10 & 4.2 & 3.0 &  19.4 & 11.1 \\

   $10^{-3}$& ${\rm O6}$ & 37 & 24.5 & 17.3 & 79.4 & 45.3 \\

  \hline\hline

\end{tabular}

\end{center}

\end{table*}


\subsection{Inner Gaseous disk structure}
In order to model the inner gaseous region of the disk, we
consider the OPAL opacity tables applicable for higher
temperatures \citep{1996ApJ...464..943I}. The various dynamical
quantities in the disk obtained using the opacity from the table
and their comparison with the various analytical models are shown
in Fig.~\ref{fig:radprofgas}. The variation of the opacity, used
for the present work, from the opacity tables with the midplane
temperature in the disk is shown in Fig.~\ref{fig:opactoomre}.

The midplane temperature reaches very high values of the order of
$ 10^{5}\, {\rm K} $ at $20 \rm {R_{\sun}}$ (See
Fig.~\ref{fig:radprofgas}).  The radial profile for the
temperature also shows a sudden break around $1\,{\rm AU}$. This
is related to the sudden rise in opacity around 3000 K as
demonstrated in Fig.~\ref{fig:opactoomre}. Since the surface
density and the mass density are inversely proportional to the
temperature (see
equations~\eqref{eq:1},~\eqref{eq:3},~\eqref{eq:5}
\&~\eqref{eq:7}), their radial profiles show a sudden fall at that
distance. Similar kind of sudden rise and fall can be seen also in
the analytical results of \citet{1994ApJ...427..987B} which are we
show alongside in Fig~\ref{fig:radprofgas}.

We also compare with analytical radial profiles obtained using
opacities as given by \citet{1991ApJ...375..740R}. However these
profiles just extent to the temperature range where the dust just
sublimates and the opacity drops to a small value. These models
fit very well in the outer disk region with the model described in
the present work.
\\
The high temperature upto $10^{5} K $  would imply a high
ionization fraction in the disk  close to the massive star. Thus
the ionized gas  could well couple to the large scale magnetic
fields from ambient medium which is dragged by accretion and may
give rise to collimated outflows.

 In Fig~\ref{fig:radprofgas}, the variation of the scale height is
also shown with the radial distance. Similar to radial profiles of
other dynamical quantities, we get jump also in the profile of the
scale height. This jump is quite interesting when dealing with
disks around massive stars. The sudden rise of disk height around
$1\, \rm{AU}$ would help to shield outer disk regions from the
radiation of central source. This affects the direction of
radiation and leading to some sort of anisotropy in the radiation
field.

We also investigate the radial profiles of the dynamical
quantities for higher mass stars. These profiles are shown in
Fig.~\ref{fig:hmass}. The midplane temperature profiles clearly
show that as the central mass and accretion rate increase, the
midplane temperature also increases and so the self sublimation
radius will move further out. For instance, with central mass as
$23\msun$ and accretion rate of $4\times10^{-3} \msun {\rm
yr}^{-1}$, the dust sublimates at $\sim 30\,{\rm AU}$ whereas the
dust sublimation radius is around $41\,{\rm AU}$ for a $37\msun$
star with disk accretion rate of $8.5\times10^{-3}\msun {\rm
yr}^{-1}$. These parameters are also chosen with the same argument
as used for $10 {\rm \msun}$ star. Also, the temperature in the
inner most region is much higher for massive young star with
higher mass. The mass density and the surface density profiles
also show the same trend of increment in the inner region with
increase in mass of central object.

On might question the thin disk approach to study the disk around
massive young stars. However, with this approach we find that the
disk around high mass stars are very similar to that around
cataclysmic variables. The spectral signatures of disk around
these compact objects were very well explained by this thin disk
model. The opacity values from the OPAL opacity table are used to
study the boundary layer of the white dwarfs
\citep{1998ApJ...502..730C}. They obtain a very high central
temperature near to the white dwarf but on applying the optical
depth consistently from the table, they get the same order of
magnitude of the effective temperature
$$T_{\rm eff}^{4} = \frac{8
T^{4}}{3 \tau}$$
as it is on the surface of a typical white dwarf.

In our case, the midplane temperature profile shows that the
region very near ($\sim 20\rsun$) is very hot gas of $10^{5}\,
{\rm K}$, but to compare with the stellar surface effective
temperature ($T_{\rm eff} = 22000\, {\rm K} $ for $M = 10\msun$)
one has to take into account the finite optical depth and then the
surface temperature of the disk, plotted in the same figure
matches the surface temperature of the star at the very innermost
regions. Thus the effective temperature of the disk can be
considerably lower than the midplane temperature. The flux from
the disk and the star are equated in order to compute the radial
distance where the stellar irradiation can affect the disk
surface,
\begin{equation}
\frac{L_\star}{4\pi r^{2}} = \sigma T_{\rm eff}^{4}
\end{equation}
where $L_{\star}$ is the luminosity of the star and $\sigma$ is
the Stefan-Boltzmann constant . We find for a $10\msun$, the
stellar irradiation will have a effect on the surface layers
around $\sim 30 {\rm AU}$ for $\alpha = 1$. If the $\alpha$ value
is lowered, the stellar irradiation can have affect on the surface
at even smaller radial distance of around 10 AU. However this
effect will not alter the dust sublimation radius in the disk.


\subsection{Growth of Massive Stellar Embryo}
One of the major problems in the formation of massive stars is the
large UV radiation from the protostars that exert pressure on the
matter inhibiting the infall on them. This is because as the
central mass increases also the central luminosity increases, thus
exerting a large radiation pressure which is usually thought to
halt the matter falling on the star. In order to get away with the
large radiant flux and to form more massive stars,
\cite{2004IAUS..221..141Y} lists some of the favorable conditions
such as $i)$ Reduction of $\kappa_{\rm eff}$ $ii)$ Reduction of
effective Luminosity and $iii)$ Increasing the gravitational
acceleration.
\subsubsection{Reduction of $\kappa_{\rm eff}$ and effective Luminosity}
 In earlier approaches considering the conditions for
the formation of massive stars, the matter was thought to be
accumulated by spherical accretion \citep{1974A&A....37..149K,
1987ApJ...319..850W}. Using detailed dust models and complex grain
size distribution, these models were able to put a limit on the
maximum stellar mass of $20 \msun$. In general, the dominance of
the radiation pressure over gravity of the spherical infall of
matter puts a strong constraint on the value of the opacity. In
this case, the necessary condition to form a massive star is
\begin{equation}\label{eq:10}
    \frac{\kappa_{\rm eff}L}{4\pi r^{2}} < \frac{GM_{*}}{r^{2}},
\end{equation}
(e.g.\ \citealt{2004IAUS..221..141Y,2007ARA&A..45..481Z}), which
implies an upper limit to the effective opacity,
\begin{equation}\label{eq:11}
    \kappa_{\rm eff} < 130 {\rm cm^{2}} {\rm g^{-1}}\left(\frac{M}{10
    \msun}\right)\left(\frac{L}{1000 {\rm L_{\odot}}}\right)^{-1}
\end{equation}
In the disk scenario, there would be additional contribution to
the forces mentioned in equation ~\eqref{eq:10}, such as disk gas
and ram pressure.

In the present work, we demonstrate that most of dust, due to the
self sublimation of accretion disk, is destroyed already at
distances larger that $ \sim 10\, \rm AU$ from the central source
which helps to reduce the $\kappa_{\rm eff}$.  It is evident that
the maximum value of the opacity even in the innermost region of
the disk is less than 5 ${\rm cm^{2}}{\rm g^{-1}}$ (see
Fig.~\ref{fig:opactoomre}), which is much smaller than the upper
limit given by equation~\eqref{eq:11} required for the formation
of a typical massive star. This suggests that the radiation
pressure is no longer inhibiting the accretion process onto to the
central star and thus allowing more massive stars to form. This
effect is aiding the matter to flow closer to the central object
but does not ensure that matter will be accreted on the central
star. Thus reduction in opacity is an essential requirement but it
does not guarantee growth of mass as the isotropic radiation from
the star may still stop the dust by exerting the pressure which is
dependent on the luminosity of the central object.

A sketch of our model calculations assuming a $10\msun$ central
star is shown in Fig.~\ref{fig:piccartoon}. The scaling of the
figure is done using a fiducial jet radius. The main components in
the figure are the dusty disk which is feeded with matter by the
core, the inner gaseous disk and the large scale bipolar outflows.
The region near to the dust sublimation radius shows the presence
of dust and gas and large variation of opacity is seen in this
region. Very close to the source there is large flux of UV and
visible radiation which is blocked from the midplane of outer
dusty disk by the high optical depths ($\tau \gg 1$) of the inner
gaseous species. The figure also shows the possibility of
radiation flux escaping more in the direction perpendicular to the
plane. This \textit{"flashlight effect"} was introduced by
\citet{1999ApJ...525..330Y} in order to have anisotropy in the
radiation and allow matter to accrete.

\subsubsection{Overcoming Radiation Pressure}
The reduction of effective opacity and luminosity ensures the
matter to come closer to the central star. In the region very near
to the star matter is subjected to different pressure sources, all
of them play an important role in understanding the dynamics of
disk accretion under the influence of stellar radiation pressure.
We therefore compare the contribution of different pressure
sources, considering typically a $10 \msun$ star.

The {\em radiation pressure from the star} is a function of the
radial distance from the surface of the star,
$$P_{\rm rad, \star}(r) = \frac{L_{\rm \star}}{4\pi c r^{2}}$$
\begin{equation}
= 5.9\times10^{2} {\rm erg} {\rm cm^{-2}}\left(\frac{T_{\rm
eff}}{22000 {\rm K}}\right)^{4}\left(\frac{r}{\rstar}\right)^{-2}
\end{equation}

 The {\em ram pressure from the disk} counteracts the radiation
force from star and will allow the matter to accrete. We estimate
the ram pressure $P_{\rm ram}=\rho v_{\rm r}^{2}$ using our
consistent dust and gas opacity model (see \S 4.1 and \S 4.3),
\setlength\arraycolsep{3pt}
\begin{eqnarray}\label{eq:23}
P_{\rm ram}(r) = 2.08\times10^{3} {\rm erg} {\rm
cm^{-2}}\left(\frac{\rho}{2.32\times10^{-9} {{\rm g} {\rm
cm^{-3}}}}\right)^{-1} \nonumber\\
\left(\frac{\dot{M}}{10^{-4}\msun {\rm yr}^{-1}}\right)^{2}
\left(\frac{M}{10 \msun}\right) \left(\frac{T}{5.03\times10^{4}
{\rm K}}\right)^{-1}
\nonumber\\
\left(\frac{\rstar}{6 \rsun}\right)^{-5}\left(\frac{r}{10
\rstar}\right)^{-5}
\end{eqnarray}

Figure~\ref{fig:Hpcomp} shows the radial profiles of both pressure
sources. The figure clearly depicts that the radiation pressure
from the star is much below the ram pressure in the disk which
will aid the matter to overcome radiation pressure and accrete on
the central star. In the very inner region of the disk, the disk
radiation pressure becomes comparable to the gas pressure in the
disk which may result in thermal
instability.(Figure~\ref{fig:Hpcomp})

\section{Stability of Disks}
The disks around massive stars can be unstable due to axisymmetric
gravitational instability. They can also be unstable due to
fragmentation due to rapid cooling of the disk as compared to the
dynamical time scale.(\citealt{2001ApJ...553..174G,
2003MNRAS.339.1025R, 2005ApJ...621L..69R}). We will also see that
these disks can be also unstable thermally in the very inner
region, close to the massive star.
\subsection{Gravitational Instability}
The criterion for the gravitational instability of a disk is given
by the Toomre parameter
$$Q = \frac{c_{s}\Omega}{2\pi G \Sigma}$$
with $\Omega$ as the Keplerian angular velocity and $\Sigma$ as
the surface density of the disk \citep{1964ApJ...139.1217T}. For
$Q > 1$  the disk is stable (no fragmentation), while $Q < 1$
leads to instability in the disk. Physically, the Toomre parameter
can be considered as the ratio of centrifugal force along the
radial direction to the gravitational force acting in the
direction perpendicular to the radial motion. Thus, if a local
accumulation of mass is moving in a certain orbit, then this would
lead to slowing down of orbital motion and also more gravitational
force acting downward which implies that the value Q will decrease
and eventually when $Q < 1$ the downward gravity force wins and
leads to instability generating overdense regions.
\symbolfootnote[3]{The Toomre criteria can be written as a product
of two ratios - $\frac{E_{thermal}}{E_{grav}}$
$\frac{E_{rot}}{E_{grav}} < 1$. If only one of the ratios is less
than unity, that will not guarantee that fragmentation will
occur.(see \S 5.2)}

 It has been known from simulations
\citep{2007ApJ...656..959K,Krumholz2009} and observations in case
of G 192.16-3.82 (\citealt{2001Sci...292.1513S}, considering large
observational errors) that disks in massive stars are unstable as
they may have Q value low and sometimes $< 1$ for some radial
extent.

The Toomre parameter decreases with the radial distance, and at
some radial distance and the disk becomes gravitational instable.
Figure~\ref{fig:opactoomre} shows the radial behavior of the
Toomre parameter in case of typical disk parameters ($\dot{M}$ =
$4.2\times10^{-4}$ $\msun$ $\rm{yr}^{-1}$ , stellar mass $M = 10
\msun$ and $\alpha \sim 1 $). It is clear that such a disk is
unstable as $Q < 1$ after $\sim 100\, {\rm AU}$.

%
\subsection {Fragmentation of Disks}

Disks which are stable for axisymmetric gravitational instability
may not be necessarily stable for fragmentation which leads to
formation of bound objects. Gravitational instability sets an
upper limit of the sound speed where as the analysis for
fragmentation and cooling time sets a lower limit on the speed of
sound in the disk\citep{2005ApJ...621L..69R}.

Numerical simulations \citep{2001ApJ...553..174G,
2003MNRAS.339.1025R} and analytical solutions
\citep{2005ApJ...621L..69R} suggest that for a disk to avoid
fragmentation, the cooling time scale should be larger than the
dynamical timescale ($t_{dyn} \sim {\Omega}^{-1}$). The threshold
relation of the cooling time can be obtained, $\Omega t_{cool} <
\zeta$, where the factor $\zeta \approx 3$, as obtained from
simulations using constant $t_{cool}$ \citep{2001ApJ...553..174G,
2003MNRAS.339.1025R}. For the present purpose, we apply cooling
time as given by \cite{2005ApJ...621L..69R},
\begin{equation}\label{eq:tcool}
t_{cool} \approx \frac{\Sigma c_s^{2}}{\gamma -
1}\frac{f(\tau)}{2\sigma T^{4}}
\end{equation}
where $f(\tau) = \tau + \frac{1}{\tau}$ describes the efficiency
of cooling and depends on the optical depth $\tau$. In the above
equation ~\eqref{eq:tcool}, $\gamma$ is the adiabatic index and
factor $\gamma -1$ considered to be of the order of
unity\citep{2005ApJ...621L..69R}.

The variation of the cooling time, as given by
eq.~\eqref{eq:tcool}, with radial distance is shown in
Fig.~\ref{fig:opactoomre} for different values of $\alpha$. These
curves are made applying the Bell \& Lin opacity power laws. The
\textit{solid line} in the figure for cooling time determines the
threshold for fragmentation to set in. The relation suggests that
a value $\alpha \sim 1$ would lead to a disk which may fragment
completely. However, the lower values of $\alpha \sim 0.1,0.01$
would enable a stable disk against fragmentation. The value
$\alpha \sim 0.1$ is consistent with that obtained by
observational modelling of ionized disks
\citep{2007MNRAS.376.1740K}.

\subsection{Thermal Instability}
Thin accretion disks around black holes which are supported by
radiation pressure (region A, See \S 2) were found to be thermally
instable \citep{1976MNRAS.175..613S}. The physical reason for such
kind of instability is inefficient cooling in the disk as compared
to the viscous heating. This leads to overheating which causes
expansion which in turn overheats the disk eventually leading to a
thermal runaway. There were ideas of application of slim disk
model to this region, which is a stable branch in the S-shaped
$\dot{M}- \Sigma$ curve, because such a model prevents the
radiation to escape from the disk and allowing the flux to be
advected along with the flow of matter
\citep{1988ApJ...332..646A}.

In the modelling of the disk around massive star, we see that the
radiation pressure becomes comparable to the gas pressure in the
very innermost region of the disk $\leq 4 R_{\star}$ for a $10 \rm
{\msun}$ star. Here the assumption that $ P_{\rm gas} > P_{\rm
rad} $ no longer holds, and we stop our iterations at this radii.
However, the application of concept of slim disk in protostellar
disks for efficient cooling is not really probable. This is
because of the fact that advected flux should get rid in some
manner, it is viable in disks around black holes as the flux can
be advected into the black holes, however in case of disks around
stars this flux will heat up the star, which may be not physical.
\citet{2007ApJ...662.1052T} applied the concept of photon bubble
instabilities to the young massive stellar environments, which
could be an efficient way of cooling the innermost region of
massive star forming region as well.

\subsection{How big is $\alpha$?}

The value of $\alpha$ parameterizes the viscosity in the disk. In
the present case, we see that higher value of $\alpha$ $\sim 1$
ensures that the disk within $\sim 100 {\rm AU}$ is stable to
axisymmetric gravitational instability, though for this high value
of $\alpha$ , the disk completely would fragment.

For $\alpha$ $\sim 0.1$ the disk would become stable to
fragmentation. In this case, the part of the disk stable to
axisymmetric gravitational instability reduces to $\leq 80-90 {\rm
AU}$

For $\alpha = 0.01$, the disk would again be stable to
fragmentation, but now most part of the disk is subjected
gravitational instability. The effect of lowering the value
$\alpha$ on the Toomre parameter can be seen in
Fig.~\ref{fig:opactoomre}

The value of $\alpha$ is usually chosen as a parameter in disk
modelling for different purposes. In case of low mass stars, the
range of $\alpha$ values accepted is around 0.001-0.01, which is
believed to be produced by magneto-rotational instability
\citep{1998RvMP...70....1B}. In case of massive stars, the value
of $\alpha$ can be higher by one order of magnitude and the
physics involved to generate this high $\alpha$ value could be
different from that seen in low stellar mass case. (for e.g.
Gravitational instability)

A high  $\alpha$ would imply that matter flows radially with high
speed near to local sound speed and there could be a possibility
of accretion shocks produced to make the radial speed of matter
subsonic. These accretion shocks have been proposed for the close
circumstellar environments of high mass protostar such as CRL 2136
with the $H_{2}O$ maser \citep{2004A&A...414..289M}.

The other way of treating viscosity in the disk would be by
assuming a radial profile, $\alpha(r)$, such that its value may be
high ($\sim 1$) in the outer region and lower($\sim 0.1$) in the
inner region of the disk. Such a radial profile would then lead to
fragmentation resulting in formation of bound objects influencing
the structure in the outer part of the disk around most massive
star formed such that each bound object may form itself a low mass
or even high mass star \citep{Krumholz2009}. However, the inner
ionized disk can be still stable and can launch large scale
bipolar outflows.

In Fig.~\ref{fig:diffalp}, variation of the dynamical quantities
obtained from the present model for a $10\msun$ star with two
different values of $\alpha$ is shown. Apart from changing the
$\alpha$ value we also change the value of mass accretion rate
according to the figure ~\ref{fig:mdotalpha}. There is no
considerable change in midplane temperature profile, which is due
to the fact that by changing two parameters also bring change in
the opacity and the cumulative effect cancels the variation in
temperature \symbolfootnote[4]{Note that $\Sigma$ is inversely
proportional to $\alpha$ and decreasing the $\alpha$ will increase
density and thus opacity}. However, this does not happen for mass
density as the dependence on the above variations is different.
Since the height depends only on the temperature, the height
profile also does not show any variation.



%


%

\section{Implications for Observations}

Early stages of massive star formation are difficult to observe,
as the whole process is enclosed in an envelope of dust. Also
these high mass star forming regions are at a distance of few kilo
parsecs which makes the observation more difficult due to limited
resolution of the present day telescopes.

One of the essential findings of the present work is the high
temperature of the order of $\sim 10^{5}\, {\rm K}$ in the inner
disk. At such a high temperature, the opacity is mostly dominated
by electron scattering and there is a pool of highly energetic
electrons present close to the massive star. These electrons might
be responsible for soft X-ray emissions from massive young stars.

There have been observations of X-rays in some of the massive
young sources in the W3 complex \citep{2002ApJ...579L..95H} and
also in GGD 27 \citep{2009ApJ...690..850P}. In the case of GGD 27,
observed hard X-rays emissions have energies of the order of 2-10
{\rm KeV}. The temperatures obtained from present work indicate
emissions of soft X-rays, however the soft X-rays emissions have a
large extinction in massive star forming regions. There might be
possibility that these less energetic electrons are accelerated by
magnetic fields giving rise to hard X-ray emission and also the
hard X-ray emission may come from shocks produced when the
accreted matter falls on the surface of star
\citep{2007IAUS..243...23M}.

Amidst all the difficulties, observers were successful to resolve
a fattened structure having length scales of $\sim 1000\, {\rm
AU}$. The best candidate of showing presence of a rotating
Keplerian disk is IRAS 21026 + 4104 \citep{2005A&A...434.1039C}
which is predicted to be early B star (Mass = 7-12 $\msun$). They
estimate the radial profile of mean kinetic temperature for the
Keplerian disk and mass density profile using different lines.
They obtain the best fit for temperature profile as $T(R) \propto
R^{-q}$ where q is 0.57-0.75 and that of mass density $ \rho(R) =
R^{-2.1}$ for $HCO^{+}$ (1-0) line. The values of the powers
obtained are very close to the best fit of these profiles from the
present model (Fig.~\ref{fig:radprofdust}).
\citet{2006ApJ...637L.129S} also were able to resolve a rotating
structure in an embedded 8-10 $\msun$ massive star AFGL 490. They
also obtain a Keplerian profile for the rotational velocity and
the best fit radial power law for the surface density was $\Sigma
\propto R^{-1.5}$, which is close to the value obtained from the
present model ($\Sigma \propto R^{-1.1}$). In case of low mass
stars, usually the density in the disk is fitted using the
gaussian as given by
\citep{2003ApJ...598.1079W,2008ApJ...674L.101W}
\begin{equation}\label{eq:18}
    \rho_{\rm disk} = \rho_{0}\left({\frac{R_{*}}{r}}\right)^{a}
    exp\left\{{\frac{-1}{2}\left[{\frac{z}{h(r)}}\right]^{2}}\right\}
\end{equation}
where the scale height has the radial dependence of the form
\begin{equation}\label{eq:19}
    h(r) = h_{0}\left({\frac{r}{R_{*}}}\right)^{b}
\end{equation}
The best fit obtained for the case of disks in "Butterfly" star
was a=2.35, b=1.28 \citep{2008ApJ...674L.101W}. These results are
interesting as similar radial fits to the elongated continuum are
obtained even for the recent observational results of young high
mass source IRAS 18151-1208 (Fallscheer private communication).
Also similar fits are obtained from the present model.


\section{Conclusion}
In this paper we have presented - for the first time - global
models for accretion disks around massive young stars.

We have solved the thin disk equations taking into account the
proper opacities for dust and gas.
In particular we consider Rosseland mean dust opacities given by
\cite{1991ApJ...375..740R} and the gas opacities from the OPAL
opacity tables \citep{1996ApJ...464..943I}.

This enables us to provide the dynamical quantities of disk
accretion from the very inner part at radii of 0.1\,AU to the
outer region of the disk at about 100\,AU.
At the same time this provides a theoretical link between outer
disk which is accessible in principle by observations and the
inner disk which is not yet possible to resolve observationally.

Our main results can be summarized as follows.
\begin{itemize}

\item[1.] For typical stellar masses and accretion rates we find
very high midplane temperatures of the order of $10^5\, {\rm K}$
for radii less than $\sim 0.1 {\rm AU}$

\item[2.] Due to the high disk temperatures the dust sublimates
already at distances which are about a factor three larger than
caused by the stellar irradiation. This {\em disk
self-sublimation} lowers the disk opacities considerably and
allows for disk accretion in the stellar radiation field.

\item[3.] We estimate the stability of these disk by the Toomre
criterion and find that our thin disks around e.g. a $10\,\msun$
star becomes gravitationally unstable beyond $100\,$AU. We also
see for $\alpha$ values close to 1, the disk would fragment
completely whereas for $\alpha \sim 0.1$ the disk could remain
stable to fragmentation. We also discuss the effect on the
dynamics and stability of the disk model with the variation of
$\alpha$.

\item[4.] For the given disk and stellar parameters and disk
opacities we find that the stellar radiation pressure is
negligible against the disk ram pressure and gas pressure and
therefore cannot hinder accretion towards a massive young star.

\end{itemize}

Considering the high disk temperatures and the rather large disk
viscosity parameter, disk around massive young stars seem to be
intrinsically different from the low mass equivalents in
particular to the different form of disk opacity

We also find that for higher mass stars with high accretion rates the dust sublimates
further away and obtain much higher temperature for the same.
The presence of the optically thick gaseous component and anisotropy of radiation
prevents the matter to be unaffected by the radiation from the star.




\clearpage
\begin{figure*}
\centering
\includegraphics[width=15.5cm]{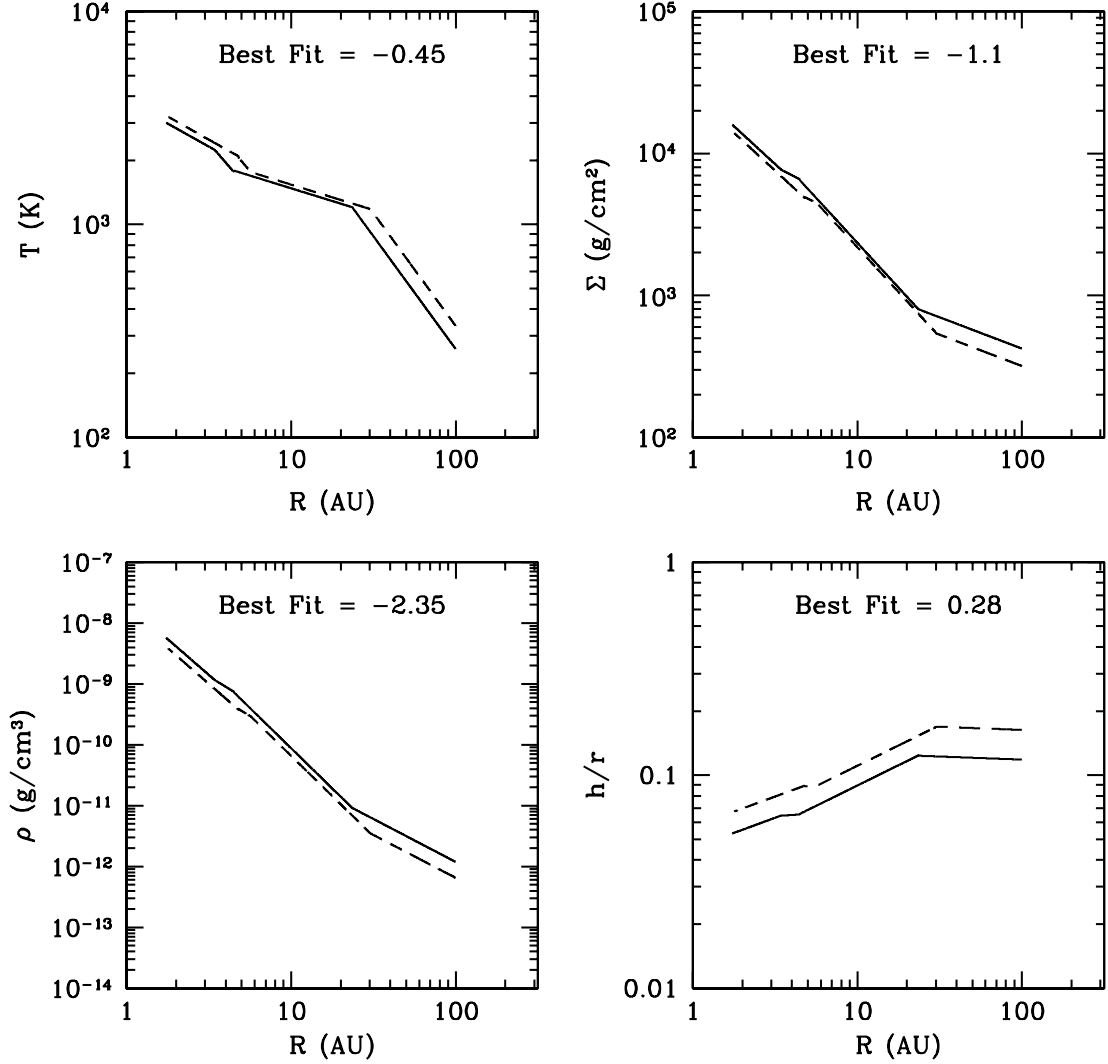}
\caption{Radial profiles of midplane temperature (T), mass density
($\rho$), surface density ($\Sigma$) and the scale height ratio ($
h/r $) for the outer dust dominated disk. The \textit{dashed line}
represents the tabulated values (see Table~\ref{tab2}), the
\textit{solid line} is obtained by using the
\citet{1991ApJ...375..740R} opacity power laws. These plots are
for typical $\dot{M}$ = $4.2\times10^{-4}$ $\msun$ $\rm{yr}^{-1}$
, stellar mass $M = 10 \msun$ and $\alpha \sim 1 $. The radial
index of the best fitted linear plot are mentioned in each
subplot.} \label{fig:radprofdust}
\end{figure*}
\clearpage

\begin{figure*}
\centering
\includegraphics[width=15.5cm]{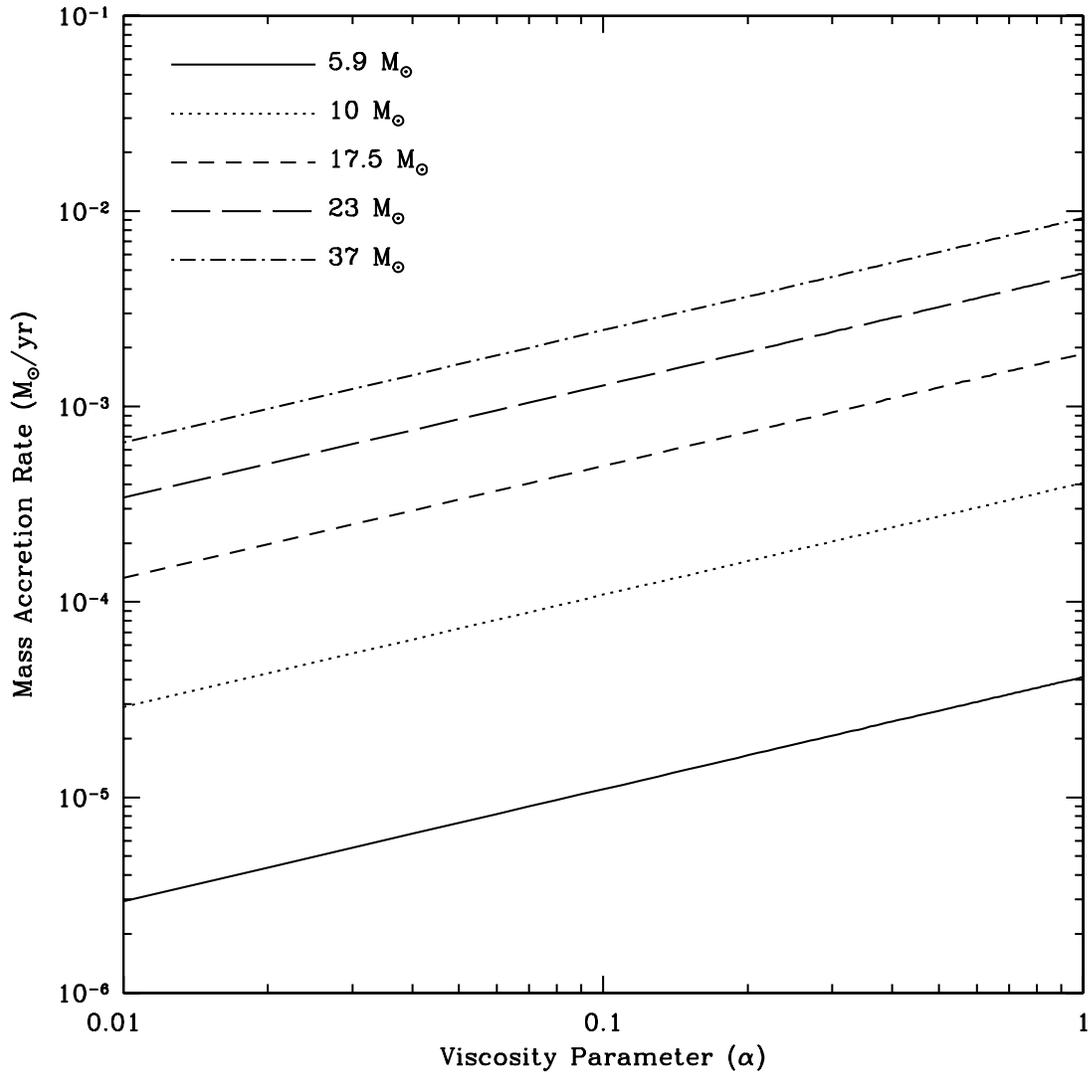}
\caption{This plot shows the variation of mass accretion rate with
the viscosity parameter so that the dust sublimation temperature
due to viscous heating is around three times that of heating from
star. The various lines are for different spectral type of stars
(luminosity values obtained from \citealt{1992adps.book.....L}).
These plots indicate that mass accretion rate is a weak function
of $\alpha$.} \label{fig:mdotalpha}
\end{figure*}
\clearpage

\begin{figure*}
\centering
\includegraphics[width=14.5cm]{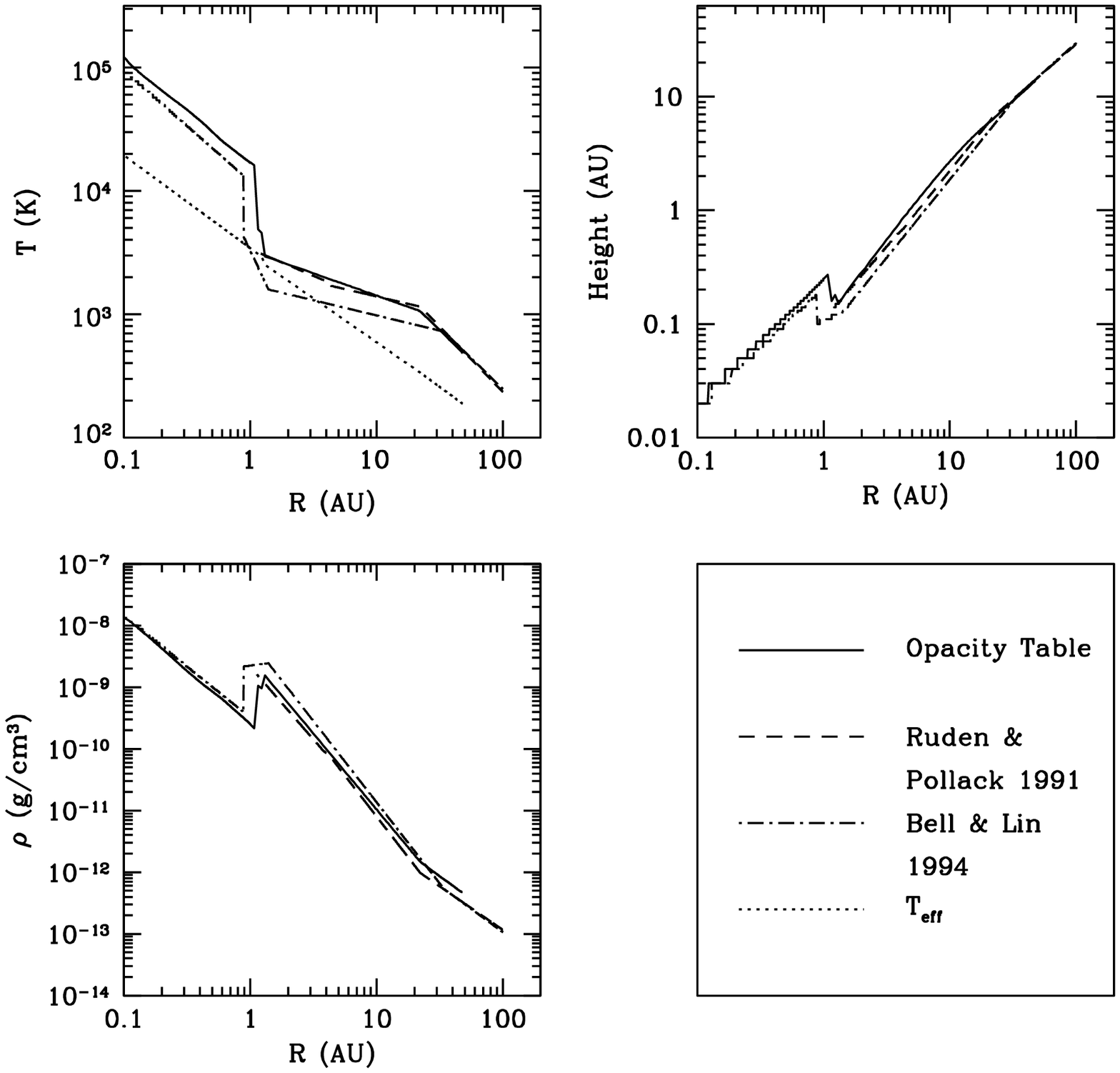}
\caption{Radial profiles of temperature (T), mass density
($\rho$), scale height of the disk extended to the inner region.
The \textit{solid line} represents the values from opacity table
\citep{2005ApJ...623..585F} and OPAL \citep{1996ApJ...464..943I} ,
the \textit{dashed line} is using the opacity power laws by
\cite{1991ApJ...375..740R} and the \textit{dot-dashed line} are
the values for Bell and Lin opacity power laws. The \textit{dotted
line} shown in temperature profile represents the $T_{\rm eff}$
profile with radius whereas the other lines are for the midplane
temperature. These plots are for typical $\dot{M}$ =
$4.2\times10^{-4}$ $\msun$ $\rm{yr}^{-1}$, stellar mass $M = 10
\msun$ and $\alpha \sim 1 $} \label{fig:radprofgas}
\end{figure*}
\clearpage

\begin{figure*}
\centering
 \includegraphics[width=13.5cm]{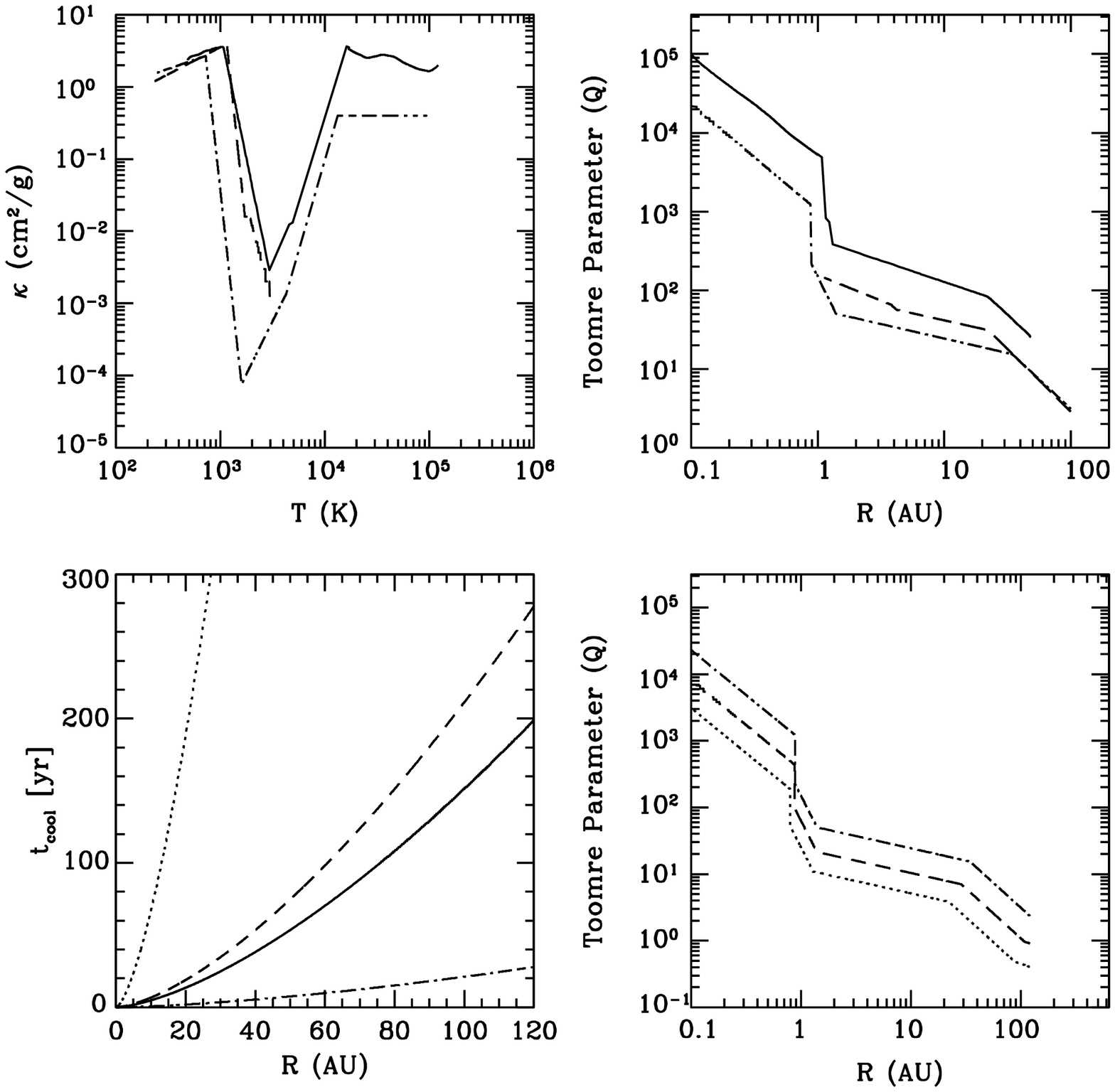}
\caption{The opacity variation with temperature ($\textit{top left
panel}$), and the variation of Toomre parameter Q with the radial
distance ($\textit{top right panel}$). The \textit{solid line}
represents the values from opacity table
\citep{2005ApJ...623..585F} and OPAL \citep{1996ApJ...464..943I},
the \textit{dashed line} is using the opacity power laws by
\cite{1991ApJ...375..740R} and the \textit{dot-dashed line} are
the values for Bell and Lin opacity power laws. The
$\textit{bottom panels}$ shows how the variation of $\alpha$
values affect the cooling time and Toomre parameter, $\alpha =
0.01$ is denoted by \textit{dotted line}, where as $\alpha = 0.1$
is \textit{dashed line} and $\alpha=1.0$ is by the
\textit{dot-dashed line}. \textit{Solid line} in the curve
represents the threshold condition for the fragmentation to set
in.} \label{fig:opactoomre}
\end{figure*}
\clearpage

\begin{figure*}
\centering
\includegraphics[width=14.5cm]{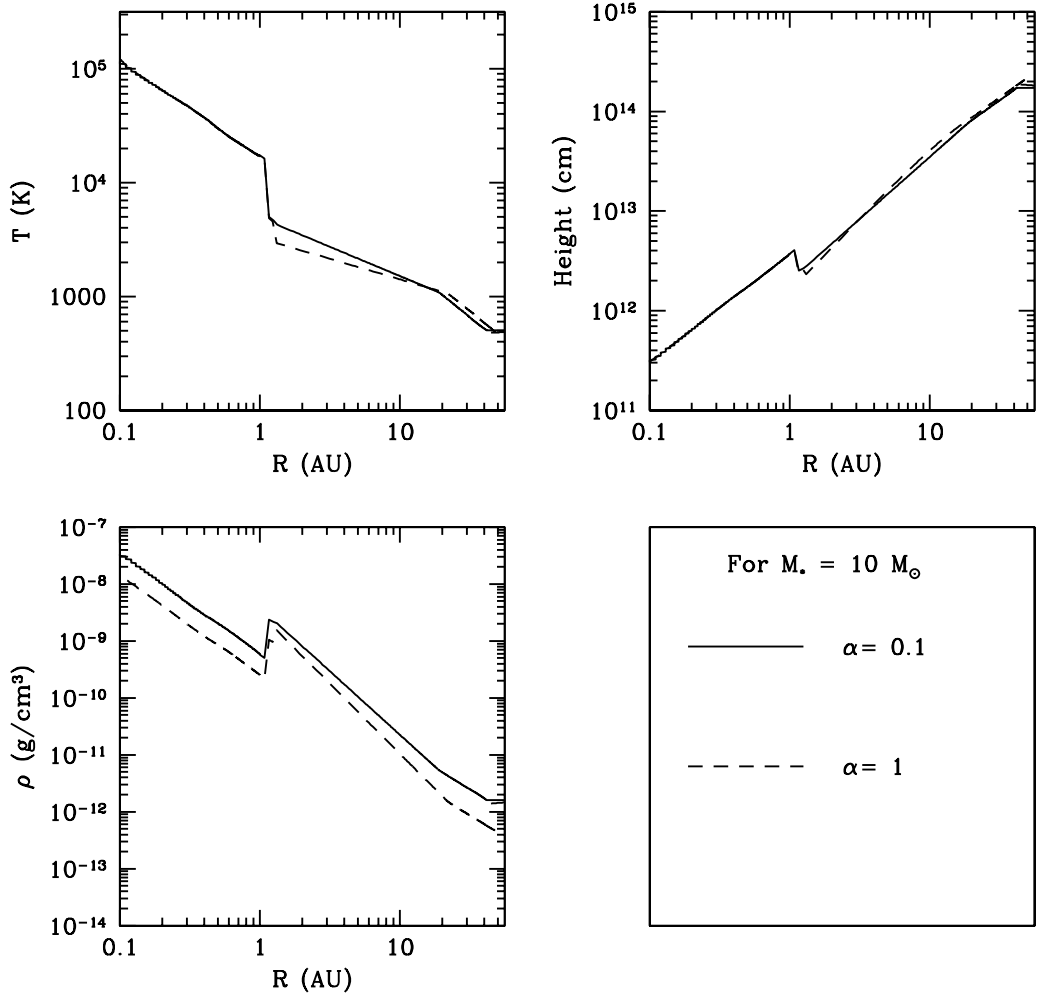}
\caption{Dynamical quantities for two different $\alpha$ values.
The \textit{solid line} represents plot with $\alpha = 0.1$ and
$\dot{M} = 10^{-4}$ where as \textit{dashed line} is for $\alpha =
1$ and $\dot{M} = 4.2\times10^{-4}$} \label{fig:diffalp}
\end{figure*}
\clearpage

\begin{figure*}
\centering
\includegraphics[width = 15.5cm]{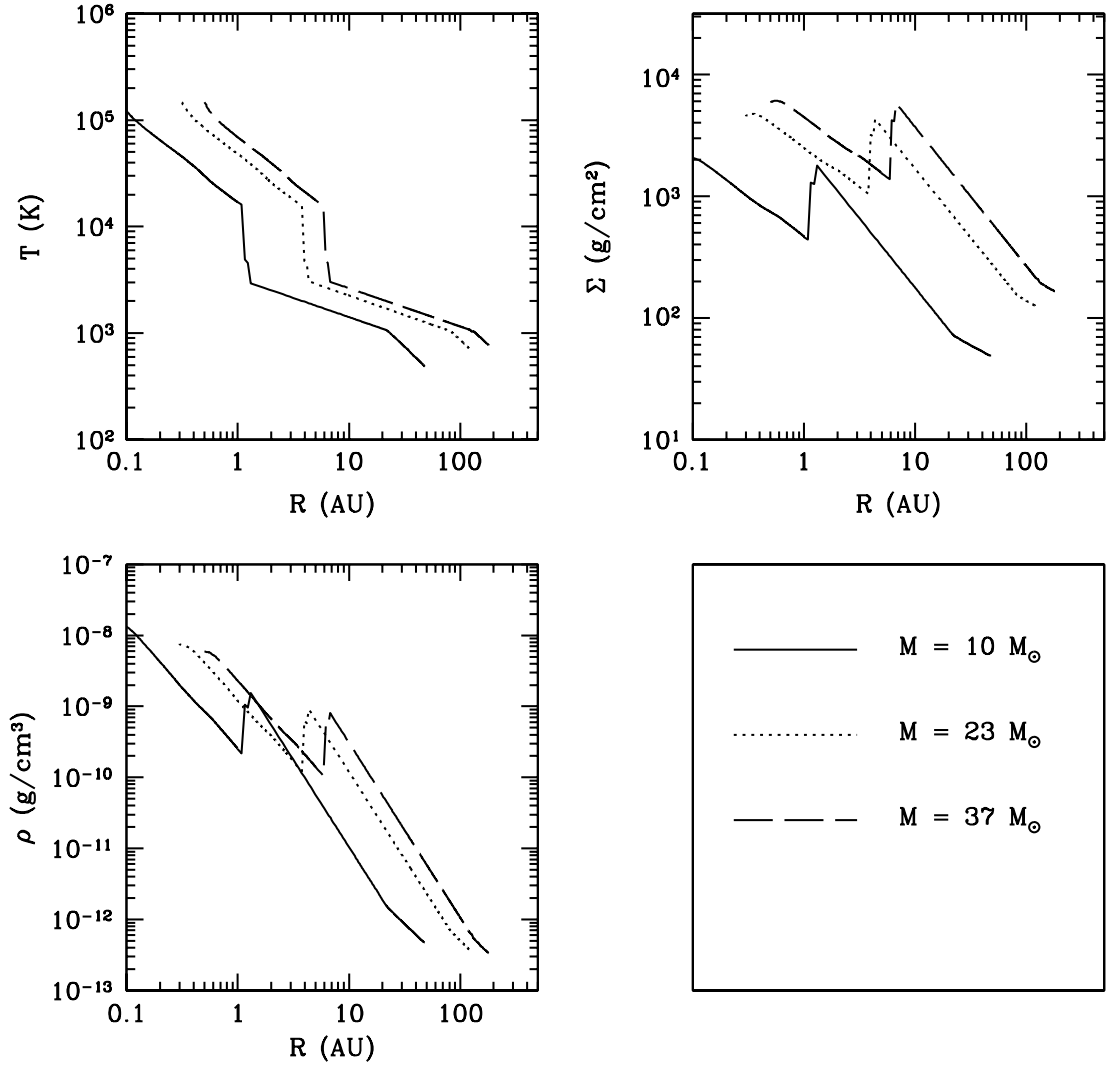}
\caption{Radial profiles of temperature (T), mass density
($\rho$), surface density ($\Sigma$) of the disk extended to the
inner region. The \textit{solid line} corresponds to $M =
10\msun$,  the \textit{dotted line} for $M = 23\msun$ and the
\textit{dashed line} is for central stellar mass $M =
37\msun$.}\label{fig:hmass}
\end{figure*}
\clearpage

\begin{figure*}
\centering
\includegraphics[width=15.5cm]{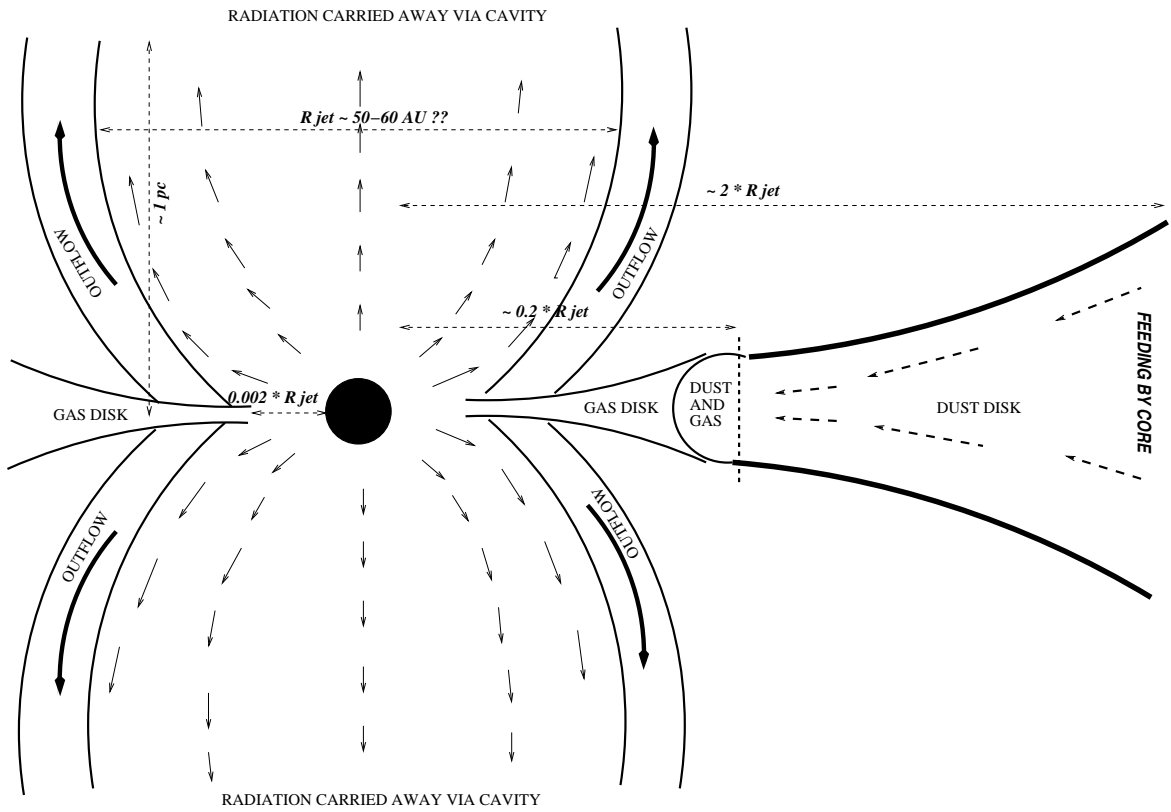}
\caption{Pictorial representation of the inner region of the
massive star.} \label{fig:piccartoon}
\end{figure*}
\clearpage

\begin{figure*}
\centering
\includegraphics[width=15.5cm]{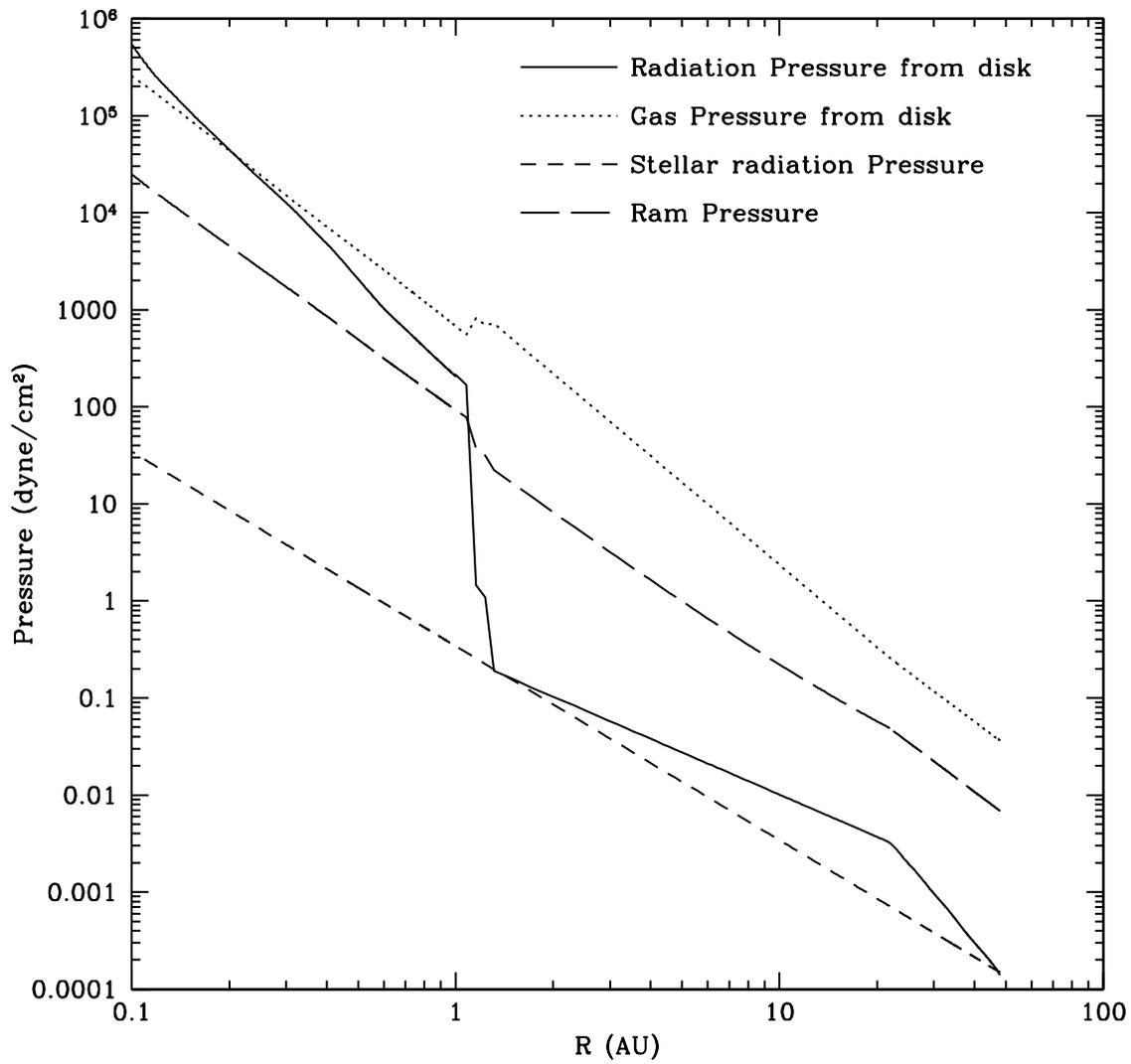}
\caption{The above figure compares the pressure due to different
sources in the circumstellar environment for a $10\,\rm {\msun}$ star}. \label{fig:Hpcomp}
\end{figure*}
\clearpage

\begin{acknowledgements}
We acknowledge the Klaus Tschira Stiftung for funding this work
carried out at Max Planck Institute of Astronomy, Heidelberg and
also convey our thanks to the Heidelberg Graduate school of
Fundamental Physics (HGSFP). We thank John Bally for his
enlightening comments on this work and also the anonymous referee
for his/her valuable comments concerning the stability analysis of
the disk. We also like to thank S. Wolf, H. Linz, C. Dullemond and
D. Semenov for their helpful comments and useful suggestions for
this work. H.B.~acknowledges financial support by the
Emmy-Noether-Program of the Deutsche Forschungsgemeinschaft (DFG,
grant BE2578). B.V. also thanks Surhud More and Mario Gennaro for
their apt suggestions.

\end{acknowledgements}




\end{document}